\documentclass[onecolumn,preprintnumbers,amsmath,amssymb]{revtex4}

\begin{document}


\title{White dwarfs as test objects of Lorentz violations}
\author{Abel Camacho}
\email{acq@xanum.uam.mx} \affiliation{Departamento de F\'{\i}sica,
Universidad Aut\'onoma Metropolitana--Iztapalapa\\
Apartado Postal 55--534, C.P. 09340, M\'exico, D.F., M\'exico.}

\date{\today}

\begin{abstract}
In the present work the thermodynamical properties of bosonic and
fermionic gases are analyzed under the condition that a modified
dispersion relation is present. This last condition implies a
breakdown of Lorentz symmetry. The implications upon the
condensation temperature will be studied, as well, as upon other
thermodynamical variables such as specific heat, entropy, etc.
Moreover, it will be argued that those cases entailing a violation
of time reversal symmetry of the motion equations could lead to
problems with the concept of entropy. Concerning the fermionic
case it will be shown that Fermi temperature suffers a
modification due to the breakdown of Lorentz symmetry. The results
will be applied to white dwarfs and the consequences upon the
Chandrasekhar mass--radius relation will be shown. The possibility
of resorting to white dwarfs for the testing of modified
dispersion relations is also addressed. It will be shown that the
comparison of the current observations against the predictions of
our model allows us to discard some values of one of the
parameters appearing in the modifications of the dispersion
relation.

\end{abstract}

\maketitle
\section{Introduction}
A long--standing puzzle in modern physics concerns the issue of a
possible quantization of the gravitational field. Some of the
current efforts in this direction entail, unavoidably, the
breakdown of Lorentz symmetry \cite{[1], [2], [3]}. In the bedrock
of modern physics lies Lorentz symmetry, and in consequence it has
been subjected to some of the highest precision tests in Physics
\cite{[4], [5], [6], [7], [8]}. Clearly, the theoretical analysis
contained in \cite{[4]} does not contemplate the study of the
consequences of a modified dispersion relation, since at that time
this idea had not yet appeared, they rather explore other
possibilities. For instance, Mansouri and Sexl consider an ether
model which maintains absolute simultaneity and elicit the
physical conclusions of this premise.

 It has to be mentioned that up to now there is no experimental evidence
 purporting a possible violation of this symmetry. At this point we have to be more careful with
 our language because the phrase violation of Lorentz symmetry could mean many things, since this
 feature embodies several characteristics, for instance, Local
 Lorentz Invariance, or Local Position Invariance \cite{[8]}. In
 our case the meaning of the phrase Lorentz violation will take a
 very precise expression. Indeed, one of the possible ways in which Lorentz symmetry could be
violated is related to the modification of the dispersion
relation. This feature emerges in some models, for instance,
non--critical string theory, non--commutative geometry, and
canonical gravity \cite{[7]}, that try to quantize the
gravitational field. In other words, we may find in the extant
attempts to quantize gravity some models that predict new physics,
since they involve a region in which one of the fundamental
symmetries of modern physics becomes only an approximation. It is
needless to say that this possibility has spurred lot of work in
this direction, but as we will point out below, more work is
required in order, either to understand better the theoretical
background of this kind of effects or to propose experiments which
could detect, or at least, put bounds upon the corresponding
parameters emerging from these models.

On the other hand, the possibility of having Lorentz symmetry only
as an approximation appears also in relation with other scenarios.
Indeed, in the phenomenological realm it has been suggested an
energy dependent speed of light as a possible solution to the GZK
paradox \cite{[9]}, i.e., the observation of ultra--high energy
cosmic ray above the expected GZK threshold for interaction of
such cosmic rays with the cosmic microwave background \cite{[9],
[10]}. In other words, there is evidence, stemming from sources
with a very diverse origin, that Lorentz symmetry could be only an
approximate feature of nature, and therefore, the analysis of the
consequences of the breakdown of the aforementioned symmetry, and
of the possible ways in which its effects could be detected, needs
no further justification.

In the experimental quest for this kind of effects interferometry
has played a fundamental role \cite{[11], [12], [13], [14]},
though it has to be underlined that it is not the unique scenario
in which experimental proposals have been introduced. In this
direction we may mention, as an interesting case, the
modifications upon the Standard Model that a Lorentz invariance
violation could have \cite{[15]}. For instance, the fact that the
maximal attainable velocity for particles is not the speed of
light and the possible detection of this difference in speed by
the Auger experiment \cite{[16]}. Within this framework photons
and neutrinos have different maximal attainable velocities, a fact
that could be detected in the next generation of neutrino
detectors as NUBE \cite{[17]}.

The idea in the present work is to introduce a deformed dispersion
relation as a fundamental fact for the dynamics of massive bosons
and fermions. Afterwards we analyze the effects of this assumption
upon the thermodynamics of the corresponding gas. Here we must
justify why the thermodynamics could suffer modifications due to
the breakdown of Lorentz symmetry. The main reason is very simple.
Statistical Mechanics teaches us that if the relation between the
momentum ($p$) and the energy ($\epsilon$) of a particle (for the
case of $l$--space--like dimensions) satisfies the relation
$\epsilon\sim p^s$, then the relation between the pressure of the
gas, $P$, and the energy density, $u$, reads $P =su/l$
\cite{[18]}. Clearly, a deformed dispersion relation modifies the
usual functional dependence upon energy and momentum, and
therefore, the thermodynamical properties do suffer changes as an
unavoidable consequence of this kind of Lorentz violation.

Clearly, the tiny effects involved in the possible deformations of
this relation entail, unfortunately, an almost unsurmountable
experimental difficulty in the case of terrestrial experimental
proposals, and in consequence, we may wonder if this kind of
analysis could shed some light in this respect. Though this last
remark is true it also has to be underlined that white dwarfs can
be considered as an example of a fermionic gas in the highly
degenerated regime \cite{[18]}. The quantal behavior of a
fermionic gas is responsible for the emergence of the
Chandrasekhar mass--radius relationship, which embodies the
equilibrium between the pressure and the gravitational interaction
of the star. As will be shown, a deformed dispersion relation
modifies the thermodynamical parameters, among them the pressure,
and, in consequence, the relation between pressure and the mass of
the star must change. In other words, white dwarfs are
astrophysical objects that could be used as a system in the quest
for this kind effects, i.e., the present proposal is then to look
for deviations in the Chandrasekhar mass--radius relationship that
could be explained by this kind of violations of Lorentz symmetry.

The possibility in this direction is supported by the
observational data \cite{[19]} where some white dwarfs have a
radius smaller than the one deduced from a normal
electron--degenerate equation of state. Though there are
theoretical efforts which try to explain this discrepancy by
several ways, for instance, a strange--quark matter within the
white dwarf core, this is not the only possibility and, in
consequence, this discrepancy could also involve in its
explanation a change in Chandrasekhar mass--radius relationship
stemming from a violation to Lorentz symmetry. Currently some
authors consider eight candidates in which this discrepancy
appears \cite{[19]}.

\bigskip
\bigskip

\section{Deformed dispersion relations and quantal properties of gasses}
\bigskip
\bigskip

As mentioned above several quantum--gravity models predict a
modified dispersion relation \cite{[1], [2], [3]}, the one can be
characterized, phenomenologically, through corrections hinging
upon Planck's length, $l_p$

\begin{equation}
E^2 = p^2\Bigl[1 - \alpha\Bigl(El_p\Bigr)^n \Bigr] + m^2.
\label{Disprel1}
\end{equation}

Here $\alpha$ is a coefficient, whose precise value depends upon
the considered quantum--gravity model, while $n$, the lowest power
in Planck's length leading to a non--vanishing contribution, is
also model dependent. Casting (\ref{Disprel1}) in ordinary units
we have ($E_p = \sqrt{c^5\hbar/G}$ denotes Planck's energy)

\begin{equation}
E^2 = p^2c^2\Bigl[1 - \alpha\Bigl(E/E_p\Bigr)^n \Bigr] + (mc^2)^2.
\label{Disprel2}
\end{equation}

Our present analysis will be restricted to massive particles, the
case of massless particles has already been studied \cite{[20]}.
At this point we will divide our study in two parts, each one of
them associated to a particular quantal statistics.

\subsection{Bosonic Statistics}

\subsubsection{General Case}

Let us consider massive bosons, according to (\ref{Disprel2}) the
relation between energy and momentum becomes now

\begin{equation}
p = \frac{1}{c}\sqrt{\frac{E^2- m^2c^4}{1 - \alpha(E/E_p)^n}}.
\label{Momentum1}
\end{equation}

The number of microstates is given by

\begin{equation}
\Sigma = \frac{s}{(2\pi\hbar)^3}\int\int d\vec{r}d\vec{p}.
\label{Numstates1}
\end{equation}

In this last expression $s$ is a weight factor arising from the
internal structure of the particles, i.e., spin. If our gas is
inside a container of volume $V$

\begin{equation}
\Sigma = \frac{4s\pi V}{(2\pi\hbar)^3}\int p^2dp.
\label{Numstates2}
\end{equation}

Casting the number of particles in terms of an integral of the
energy

\begin{equation}
\Sigma = \frac{4s\pi V}{(2\pi\hbar)^3}\int_0^{\infty}\frac{1}{c^3}
\sqrt{\frac{E^2- m^2c^4}{1 -
\alpha(E/E_p)^n}}\Bigl\{\frac{E+\alpha[(n-1)E^2
-nm^2c^4](E^{n-1}/E^{n}_p)}{[1-\alpha(E/E_p)^n]^2}\Bigr\}dE.
\label{Numstates3}
\end{equation}

The density of states per energy unit is easily calculated

\begin{equation}
\Omega(E) = \frac{4s\pi V}{(2c\pi\hbar)^3}\sqrt{\frac{E^2-
m^2c^4}{1 - \alpha(E/E_p)^n}}\Bigl\{\frac{E+\alpha[(n-1)E^2
-nm^2c^4](E^{n-1}/E^{n}_p)}{[1-\alpha(E/E_p)^n]^2}\Bigr\}.
\label{Denstates1}
\end{equation}

If in (\ref{Denstates1}) we set $\alpha =0$, $s=2$, and $m=0$,
then we recover the density of states for photons \cite{[18]}

\begin{equation}
\Omega(E) = \frac{8\pi V}{(2c\pi\hbar)^3}E^2. \label{Denstates2}
\end{equation}

The average number of particles reads

\begin{equation}
N= N_0 + N_e,\label{Avernumber1}
\end{equation}

\begin{equation}
N_0 = \frac{s}{1-\lambda\exp(-mc^2/\kappa T)},\label{Avernumber2}
\end{equation}

\begin{equation}
N_e =
s\int_{mc^2}^{\infty}\frac{\Omega(E)}{\lambda^{-1}e^{E/\kappa T}
-1}dE.\label{Avernumber13}
\end{equation}

Here $N_0$ denotes the number of particles in the ground state,
whereas $N_e$ is the number of particles in the excited states.
Additionally, $\kappa$ and $\lambda$ are Boltzmann's constant and
the fugacity \cite{[18]}, respectively. Performing the integration
we obtain

\begin{eqnarray}
N_e =
 \frac{4s\pi V}{(2c\pi\hbar)^3}(\kappa T)^3\Bigl\{\Gamma(3)g_{(3)}(\lambda\exp(-mc^2/\kappa T))+
 \alpha(n+3/2)(\kappa T/E_p)^n
\Bigl[\Gamma(3+n)g_{(3+n)}(\lambda\exp(-mc^2/\kappa T))
\nonumber\\
+\frac{n+1}{2}\Gamma(1+n)(mc^2/\kappa
T)^2g_{(1+n)}(\lambda\exp(-mc^2/\kappa
T))\Bigr]\Bigr\}.\label{Avernumber3}
\end{eqnarray}

In (\ref{Avernumber3}) we have that $g_{(n)}(z)$ are the
Bose--Einstein functions \cite{[18]}, $\Gamma(n)$ the
Gamma--functions \cite{[21]}. Additionally, $g_{(n)}(z)$ is a
monotonically increasing function, and, if $n>1$, then it is
bounded $\forall z\in [0, 1]$. In other words, $N_e$ has a maximum
defined by the fulfillment of the condition

\begin{eqnarray}
\lambda\exp(-mc^2/\kappa T) = 1.\label{Max1}
\end{eqnarray}

This entails that the number of particles that can be located in
the excited states has a maximum

\begin{eqnarray}
N^{(max-n)}_e =
 \frac{4s\pi V}{(2c\pi\hbar)^3}(\kappa T)^3\Bigl\{2\xi(3)+
 \alpha(n+3/2)(T/T_p)^n
\Bigl[(n+2)!\xi(3+n) + \frac{(n+1)!}{2}
\nonumber\\
\times (mc^2/\kappa T)^2\xi(n+1)\Bigr]\Bigr\}.\label{Avernumber4}
\end{eqnarray}

We have introduced the definition of Planck's temperature $T_p =
E_p/\kappa$, and the Riemann Zeta function $\xi(z)$ \cite{[22]}.
If we impose the condition $\alpha =0$ then

\begin{eqnarray}
N^{(max)}_e =
 \frac{8s\pi V}{(2c\pi\hbar)^3}(\kappa T)^3\xi(3).\label{Avernumber5}
\end{eqnarray}

In other words, we recover a fact already known, see expression
(13) in \cite{[23]}. If $\alpha
>0$ then the maximum number of particles in the excited states
grows, whereas, if $\alpha <0$, this quantity diminishes. Hence,
in the former case the number of particles in the ground state,
$N_0$, diminishes, i.e., the Bose--Einstein condensation "slows
down", for the latter case the condensation "speeds up". In the
context of Bose--Einstein condensation the breakdown of Lorentz
symmetry appears as a modification of the number of particles that
can be located in the excited states. In order to evaluate this
change notice that

\begin{eqnarray}
\vert N^{(max-n)}_e- N^{(max)}_e \vert/N^{(max)}_e=
 \vert\alpha\vert\frac{n+3/2}{\xi(3)}(T/T_p)^n
\Bigl[(n+2)!\xi(3+n) + \frac{(n+1)!}{2}
\nonumber\\
\times (mc^2/\kappa T)^2\xi(n+1)\Bigr]\Bigr\} .\label{Numchange1}
\end{eqnarray}

Unfortunately the effect depends upon the ratio $T/T_p$, hence

\begin{eqnarray}
\vert N^{(max-n)}_e- N^{(max)}_e\vert /N^{(max)}_e\sim
0.1\rightarrow T\sim
\Bigl[\frac{1}{\vert\alpha\vert(n+3/2)}\Bigr]^{1/n}T_p.\label{Numchange2}
\end{eqnarray}

The required temperature is a linear function of Planck's
temperature, a fact that experimentally is a serious drawback,
since $T_p\sim 10^{32}K$, and in order to have a noticeable we
must achieve temperatures close enough to $T_p$.

Let us now address the issue of the number of particles in the
ground state. The condition for the existence of Bose--Einstein
condensation reads

\begin{eqnarray}
N > N^{(max-n)}_e.\label{Boseconde1}
\end{eqnarray}

As in the case in which Lorentz symmetry is present here a
critical temperature, $T_c$, appears

\begin{eqnarray}
N =
 \frac{4s\pi V}{(2c\pi\hbar)^3}(\kappa T_c)^3\Bigl\{2\xi(3)+
 \alpha(n+3/2)(T_c/T_p)^n
\Bigl[(n+2)!\xi(3+n) + \frac{(n+1)!}{2}
\nonumber\\
\times(mc^2/\kappa T_c)^2\xi(n+1)\Bigr]\Bigr\}.\label{Critem1}
\end{eqnarray}

Since $N = N_0 + N_e$, then $N_0/N = 1 - N_e/N$, and, according to
(\ref{Critem1}) and (\ref{Avernumber4}), we have that

\begin{eqnarray}
N_e/N =
\Bigl(\frac{T}{T_c}\Bigr)^3\frac{2\xi(3)+\alpha(n+3/2)(T/T_p)^n\Bigl[(n+2)!\xi(3+n)
+ \frac{(n+1)!}{2}(mc^2/\kappa
T)^2\xi(n+1)\Bigr]}{2\xi(3)+\alpha(n+3/2)(T_c/T_p)^n\Bigl[(n+2)!\xi(3+n)
+\frac{(n+1)!}{2}(mc^2/\kappa
T_c)^2\xi(n+1)\Bigr]}.\label{Critem2}
\end{eqnarray}

This last expression allows us to calculate the ratio $N_0/N$.
Indeed, if $T>T_c$, then $N_e/N =1$, whereas if $T<T_c$, we must
resort to (\ref{Critem2}) to evaluate $N_e/N$. Setting $\alpha =0$
we recover the usual expression \cite{[18], [23]}. The breakdown
of Lorentz symmetry does impinge upon the condensation
temperature.

Let us now analyze some other thermodynamical parameters, for
instance, the internal energy, $U$.

For our case

\begin{eqnarray}
U =\frac{smc^2}{\lambda^{-1}\exp[mc^2/\kappa T] -1} + \frac{4s\pi
V}{(2\pi c\hbar)^3}\int_{mc^2}^{\infty}\Bigl\{E^2\sqrt{E^2 -
m^2c^4} +
\alpha\frac{n+3/2}{E^n_p}E^{n+2}\sqrt{E^2-m^2c^4}\Bigr\}\frac{dE}{\lambda^{-1}\exp[E/\kappa
T] -1}.\label{Intenergy1}
\end{eqnarray}

We may cast the internal energy in the following form

\begin{eqnarray}
U = smc^2\lambda\exp\Bigl\{-mc^2/\kappa T\Bigr\}N_0 +\frac{4s\pi
V}{(2\pi c\hbar)^3}\int_{0}^{\infty}\frac{E^2\sqrt{E^2 +
m^2c^4}dE}{\lambda^{-1}\exp[\sqrt{E^2+m^2c^4}/\kappa
T]-1}\nonumber\\
+ \frac{4s\pi V}{(2\pi c\hbar)^3}\alpha(n+3/2)(\kappa
T)^4\Bigl(T/Tp\Bigr)^n\Sigma_{l=0}^{n+1}\frac{(n+1)!}{l!(n+1-l)!}\Bigl(\frac{mc^2}{\kappa
T}\Bigr)^l\Gamma(n+4-l)g_{(n+4-l)}(\lambda\exp(-mc^2/\kappa T))
 .\label{Intenergy2}
\end{eqnarray}

This shows that the internal energy is modified if Lorentz
symmetry is broken. The energy grows if $\alpha >0$, and
diminishes when $\alpha <0$. Unfortunately the effect, as pointed
out before, behaves as $(T/T_p)^n$, remember that $T_p =
E_p/\kappa$ denotes Planck's temperature..

Let us now analyze the pressure of our gas. With our assumptions
we may obtain that the pressure is given by

\begin{eqnarray}
P = \frac{4s\pi V}{(2\pi
c\hbar)^3}\Bigl\{\frac{1}{3}\int_{0}^{\infty}\frac{E^4}{\sqrt{E^2
+ m^2c^4}}\frac{dE}{\lambda^{-1}\exp[\sqrt{E^2+m^2c^4}/\kappa
T]-1}\nonumber\\
+\alpha(n+3/2)(\kappa
T)^4\Bigl(T/Tp\Bigr)^n\Sigma_{l=0}^{n}\frac{(n)!}{l!(n-l)!}\Bigl(\frac{mc^2}{\kappa
T}\Bigr)^l\Gamma(n+3-l)g_{(n+4-l)}(\lambda\exp(-mc^2/\kappa
T))\Bigr\}.\label{Pressure1}
\end{eqnarray}

It is important to mention that if $\alpha >0$, then the pressure
grows, with respect to the case in which Lorentz symmetry is
present.

This last remark allows us to interpret the breakdown of Lorentz
symmetry for massive bosons as a repulsive interaction, if $\alpha
>0$. Indeed, the presence of a repulsive interaction (among the
particles of a gas) entails the increase of the pressure, compared
against the corresponding value for an ideal gas.

Let us explain this point deeper. A fleeting glimpse at the
cluster expanssion and its relation to the virial coefficients
\cite{[18]} clearly shows that the first correction to the ideal
gas state equation expressed in terms of the virial state equation
($PV/(NKT) =
\Sigma_{l=1}^{\infty}a_l(T)\bigl(N\lambda^3/V\bigr)^{l-1}$, where
$N$ denotes the number of particles) corresponds to a virial
coefficient that can be written as a function of the potential
energy of interaction between the i--th and the j--th particle
$v_{ij}$

\begin{equation}
a_2 = -\frac{1}{\lambda}\int f_{12}d^3r_{12}.\label{2virialcoeff}
\end{equation}

Where $\exp\bigl\{-v_{12}/KT\bigr\} = 1 + f_{12}$, and $\lambda =
\sqrt{2\pi\hbar^2/mKT}$ is the thermal wavelength \cite{[18]}.

A repulsive interaction means that $v_{12}>0$, and in consequence
$f_{12}<0$, and therefore, $a_2>0$. If we introduce this condition
into the virial expression we obtain a pressure larger than that
corresponding to an ideal gas. In other words, the introduction of
a repulsive interaction among the particles comprising the gas
entails an increase of the pressure, compared to the pressure of
an ideal gas.

It is in this sense that we say that the loss of the symmetry
appears, at the bulk level, as the emergence of a repulsive
interaction (if $\alpha >0$) , and in consequence, at least in
principle, we could detect some effects stemming from loop quantum
gravity, non--commutative geometry, etc. If $\alpha$ turns out to
be negative, then the breakdown is equivalent, for massive bosons,
to an attractive interaction.

Another interesting point is related to the density of number of
states, see (\ref{Denstates1}). Clearly, if $\alpha >0$, then the
density grows, whereas, if $\alpha <0$, this parameter decreases.
This last remark allows us to state that, since the entropy is a
function of the number of microstates available to a macrostate,
then a positive $\alpha$ must yield a larger entropy (compared to
the case in which Lorentz symmetry is present), and if $\alpha$ is
negative, then we must have a smaller entropy. This assertion can
be checked recalling that the entropy $S$ satisfies the relation
\cite{[18]}

\begin{equation}
\frac{S}{N\kappa} = \frac{U +PV}{N\kappa T} - \frac{\mu}{\kappa
T}.\label{Entropy1}
\end{equation}

In this last expression $\mu$ denotes the chemical potential. Our
previous results allow us to write

\begin{eqnarray}
S = \frac{4s\pi V}{(2\pi
c\hbar)^3T}\Bigl\{\frac{4}{3}\int_{0}^{\infty}\frac{E^4}{\sqrt{E^2
+ m^2c^4}}\frac{dE}{\lambda^{-1}\exp[\sqrt{E^2+m^2c^4}/\kappa
T]-1}\nonumber\\
+ \alpha(n+3/2)(\kappa
T)^4\Bigl(T/Tp\Bigr)^n\Bigl[\Sigma_{l=0}^{n}\frac{(n)!}{l!(n-l)!}\Bigl(\frac{mc^2}{\kappa
T}\Bigr)^l\Gamma(n+3-l)g_{(n+4-l)}(\lambda\exp(-mc^2/\kappa T))
\nonumber\\
+\Sigma_{l=0}^{n+1}\frac{(n+1)!}{l!(n+1-l)!}\Bigl(\frac{mc^2}{\kappa
T}\Bigr)^l\Gamma(n+4-l)g_{(n+4-l)}(\lambda\exp(-mc^2/\kappa
T))\Bigr]\Bigr\}-N\kappa ln\lambda + \lambda mc^2\exp(-mc^2/\kappa
T)N_0/T .\label{Entropy2}
\end{eqnarray}

Indeed, if $\alpha >0$, then we obtain an entropy larger than the
corresponding value with Lorentz symmetry. The remaining
possibility, $\alpha <0$, embodies a smaller entropy. In the limit
$T\rightarrow 0$ the entropy $S$ vanishes, as happens in the usual
case \cite{[18]}.

\subsubsection{Reversible Adiabatic Processes}

Let us now resort to (\ref{Entropy2}) and consider the case $n=1$,
the entropy becomes

\begin{eqnarray}
S_{(1)} = \frac{4s\pi V}{(2\pi
c\hbar)^3T}\Bigl\{\frac{4}{3}\int_{0}^{\infty}\frac{E^4}{\sqrt{E^2
+ m^2c^4}}\frac{dE}{\lambda^{-1}\exp[\sqrt{E^2+m^2c^4}/\kappa
T]-1}\nonumber\\
+ \alpha(5/2)(\kappa
T)^4\Bigl(T/Tp\Bigr)\Bigl[3!g_{(5)}(\lambda\exp(-mc^2/\kappa T))+
2!\Bigl(\frac{mc^2}{\kappa
T}\Bigr)g_{(4)}(\lambda\exp(-mc^2/\kappa T))+\nonumber\\
4!g_{(5)}(\lambda\exp(-mc^2/\kappa T)) +
2!\Bigl(\frac{mc^2}{\kappa
T}\Bigr)3!g_{(4)}(\lambda\exp(-mc^2/\kappa T)) +
2!\Bigl(\frac{mc^2}{\kappa
T}\Bigr)^22!g_{(3)}(\lambda\exp(-mc^2/\kappa
T))\Bigr]\Bigr\}\nonumber\\
-N\kappa ln\lambda + \lambda mc^2\exp(-mc^2/\kappa T)N_0/T
.\label{Entropy3}
\end{eqnarray}

We know that the breakdown of Lorentz symmetry implies also the
violation of the CPT theorem \cite{[24]}, since it assumes the
presence of Lorentz symmetry. Therefore, we may wonder if this
violation could have some consequences at macroscopic level. For
instance, let us assume that the case $n=1$ is related to the
breakdown of time reversal invariance, i.e., it entails a theory
in which the dynamics is not time reversal invariant. Then, the
concept of reversible adiabatic process is lost, since it
necessarily requires time reversal invariant laws of motion. An
acceptable definition of entropy must satisfy eight criteria
\cite{[25]}, among them we may find the fact that the entropy must
be invariant in all reversible adiabatic processes, and increase
in any irreversible adiabatic process (the second criterion of
\cite{[25]}). Within this context consider the case in which our
bosonic gas is inside a container whose walls are adiabatic, and
now let us introduce the following condition

\begin{eqnarray}
S_{(1)} = const.\label{Reversible1}
\end{eqnarray}

It is an adiabatic process in which the entropy does not change.
Since we have assumed the violation of time reversal invariance,
then (\ref{Reversible1}) has no physical meaning, since it is
related to reversible processes, which do not exist if time
reversal invariance is lost. In other words, in those cases in
which the breakdown of Lorentz symmetry involves a violation of
time reversal invariance we will end up with a concept of entropy
which will define curves without physical meaning. Then it seems
that we face two possibilities: (i) either thermodynamics loses
its validity within this context; (ii) or those cases implying the
loss of time reversal invariance have to be discarded, since they
violate thermodynamics.

\subsection{Fermionic Statistics}

Let us now consider the case of a fermionic gas in which, once
again, we have a deformed dispersion relation as that given by
(\ref{Disprel2}). The argument mentioned above (explaining the
modifications of the state equation for a bosonic gas) is valid
here, and, in consequence, we must expect changes in the
thermodynamical variables of the gas, if Lorentz symmetry is
broken. One of these parameters is the Fermi momentum, which
defines, in the limit $T\rightarrow 0$, the border between
single--particle states empty and occupied. Indeed, in the
aforementioned limit the mean occupation numbers of
single--particle state, $<n_{E}>$, are equal to $0$ if $E>\mu_0$,
and $1$ if $E<\mu_0$, where $\mu_0$ is the chemical potential at
$T=0$ \cite{[18]}.

For our case the zero--point energy of an electronic gas (with $N$
particles in a volume $V$) becomes

\begin{eqnarray}
\epsilon_0 = 2\Sigma_{(p<P_F)}\sqrt{p^2c^2\Bigl[1 -
\alpha\Bigl(E/E_p\Bigr)^n \Bigr] + (mc^2)^2}.\label{Groundenergy1}
\end{eqnarray}

The sum has to be done considering only those states with a
momentum smaller than Fermi momentum, $P_F$, since those with a
higher momentum are empty \cite{[18]}.

In the continuum limit we obtain that

\begin{eqnarray}
\epsilon_0 = \frac{2V}{(2\pi \hbar)^3}\int_0^{P_F} 4\pi
p^2\sqrt{(pc)^2+(mc^2)^2}\Bigl\{1 -
\frac{\alpha}{2}\frac{(pc)^2}{(pc)^2+
(mc^2)^2}\Bigl(\frac{\sqrt{(pc)^2+(mc^2)^2}}{E_p}\Bigr)^n\Bigr\}dp
.\label{Groundenergy2}
\end{eqnarray}

This last expression can be cast in the following form

\begin{eqnarray}
\epsilon_0 = \frac{8\pi V}{(2\pi \hbar)^3}(m^4c^5)\Bigl\{f(x_F) -
\frac{\alpha}{2}(\frac{mc^2}{E_p})^ng(x_F)\Bigr\}
,\label{Groundenergy3}
\end{eqnarray}

\begin{eqnarray}
x_F = \frac{P_F}{mc},\label{Fermipara1}
\end{eqnarray}

\begin{eqnarray}
P_F = \hbar(\frac{3\pi^2 N}{V})^{1/3},\label{Fermipara2}
\end{eqnarray}

\begin{eqnarray}
g(x_F) = \int_0^{x_F}x^4(1+x^2)^{(n-1)/2}dx,\label{Fermifun1}
\end{eqnarray}

\begin{eqnarray}
f(x_F) = \frac{1}{16}\Bigl\{\frac{1}{4}\Bigl[\Bigl(x_F
+\sqrt{1+x^2_F}\Bigr)^4 - \Bigl(x_F +\sqrt{1+x^2_F}\Bigr)^{-4}
\Bigr] -\frac{1}{2}ln[x_F
+\sqrt{1+x^2_F}]\Bigr\}.\label{Fermifun2}
\end{eqnarray}

It is readily seen that the dependence upon $n$ of the breakdown
of Lorentz symmetry is encoded in $g(x_F)$ and in the parameter
$(\frac{mc^2}{E_p})^n$. For the sake of simplicity let us consider
the case $n=1$, then

\begin{eqnarray}
\epsilon_0 = \frac{8\pi V}{(2\pi \hbar)^3}(m^4c^5)\Bigl\{f(x_F) -
\frac{\alpha mc^2}{10E_p}x^5_F\Bigr\}.\label{Groundenergy4}
\end{eqnarray}

Since the pressure of the zero--point energy is given by $P_0 =
-\Bigl(\frac{\partial\epsilon_0}{\partial V}\Bigr)$ \cite{[18]},
we obtain for this case ($n=1$)

\begin{eqnarray}
P_0 = \frac{8\pi}{(2\pi
\hbar)^3}(m^4c^5)\Bigl\{\frac{x_F}{3}\frac{df(x_F)}{dx_F} -
f(x_F)- \frac{\alpha
mc^2}{15E_p}x^5_F\Bigr\}.\label{Groundpressure1}
\end{eqnarray}

If we consider $\alpha >0$ then, the pressure decreases with
respect to the case in which Lorentz symmetry is not broken,
whereas, if $\alpha <0$ the pressure grows. Once again, we say
that the loss of the symmetry appears, at the bulk level, as an
attractive interaction (if $\alpha
>0$). When $\alpha$ turns out to be negative, then the breakdown is
equivalent, for massive fermions, to a repulsive interaction.

\section{White Dwarfs}

At this point we may wonder if a fermionic system could provide us
with some experimental proposal which could lead us to have a more
optimistic scenario than the one appearing in connection with
bosonic gasses. Historically, Fermi statistics was first applied
in astrophysics, namely, the thermodynamic equilibrium of white
dwarfs \cite{[26], [27]}. An idealized white dwarf consists of
Helium in an almost complete state of ionization, and hence the
microscopic constituents of the star may be taken a $N$ electrons
and $N/2$ Helium nuclei at temperature in which the dynamics of
the electrons is in the relativistic limit (in other words,
$x_F>>1$ \cite{[18]}), additionally, the helium nuclei do not
contribute significantly to the dynamics of the problem. In other
words, in a first approximation we may neglect the presence of the
nuclei in the system. The temperature of the star ($T\sim 10^7 K$)
allows us to consider the electron gas in a state of almost
complete degeneracy ($T_F = 10^{10} K$).

The equilibrium configuration of a white dwarf is given by the
fact that the pressure of the gas competes with the gravitational
attraction. Clearly, (\ref{Groundpressure1}) shows us that the
pressure of the zero--point energy is modified as a consequence of
the breakdown of Lorentz symmetry, and therefore, we must expect a
change in the analysis of the equilibrium state of a white dwarf.
In other words, Chandrasekhar shows us that the equilibrium of
white dwarfs defines a relationship between the mass of the star
and its radius \cite{[27]}, and in consequence taking into account
our arguments we may seek for violations of Lorentz symmetry
looking for changes in the Chandrasekhar mass--radius
relationship.

 Let us assume that the star has a spherical configuration (of radius $R$)
and that the electron gas is uniformly distributed over the body
of the star, this restriction is easily removed, but we consider
it in our first order approximation. If an adiabatic change in $V$
takes place, then the change in the zero point energy is given by

\begin{eqnarray}
d\epsilon_0 = -4\pi R^2P_0(R)dR.\label{Energychange1}
\end{eqnarray}

The change in the gravitational potential energy is given by

\begin{eqnarray}
dE_g = \frac{GM^2}{R^2}dR.\label{Energychange2}
\end{eqnarray}

In this last expression $M$ is the mass of the star and $G$ the
gravitational constant \cite{[27]}. If the system is in
equilibrium, then the net change in its total energy, ($E_0 +
E_g$), for an infinitesimal change in its size, should be zero,
thus, for equilibrium

\begin{eqnarray}
P_0(R) = \frac{GM^2}{4\pi R^4}.\label{Equilibrium1}
\end{eqnarray}

Notice that (\ref{Groundpressure1}) entails that the pressure
depends upon the density (see (\ref{Fermipara2})), and therefore

\begin{eqnarray}
P_F = \hbar\Bigl(\frac{9\pi
N}{4R^3}\Bigr)^{1/3},\label{Fermipara3}
\end{eqnarray}

\begin{eqnarray}
x_F = \frac{\hbar}{mcR}\Bigl(\frac{9\pi
N}{4}\Bigr)^{1/3}.\label{Fermipara4}
\end{eqnarray}

Since $M\approx 2Nm_p$, where $m_p$ is the proton mass, then ($m$
is the electron mass)

\begin{eqnarray}
x_F = \frac{\hbar}{mcR}\Bigl(\frac{9\pi
M}{8m_p}\Bigr)^{1/3}.\label{Fermipara5}
\end{eqnarray}

\subsection{Case n=1}

The equilibrium condition reads

\begin{eqnarray}
\frac{GM^2}{4\pi R^4} = \frac{8\pi}{(2\pi
\hbar)^3}(m^4c^5)\Bigl\{\frac{x_F}{3}\frac{df(x_F)}{dx_F} -
f(x_F)- \frac{\alpha mc^2}{15E_p}x^5_F\Bigr\}.\label{Equilibrium1}
\end{eqnarray}

From this expression we may find the dependence of $R$ upon $M$,
the one is given by

\begin{eqnarray}
R = \Gamma\sqrt{1 - \frac{12\beta}{\tau\Gamma^4}}\Bigl\{1 -
\frac{2\alpha\omega}{5}\Bigl[1 -
\frac{12\beta}{\tau\Gamma^4}\Bigr]^{-3/2}\Bigr\}
,\label{massradius1}
\end{eqnarray}

\begin{eqnarray}
\Gamma = \frac{\hbar}{mc}\Bigl(\frac{9\pi
M}{8m_p}\Bigr)^{1/3},\label{Constant1}
\end{eqnarray}

\begin{eqnarray}
\omega = \frac{mc^2}{E_p},\label{Constant2}
\end{eqnarray}

\begin{eqnarray}
\beta = \frac{GM^2}{4\pi},\label{Constant3}
\end{eqnarray}

\begin{eqnarray}
\tau = \frac{8\pi}{(2\pi\hbar)^3}(m^4c^5).\label{Constant4}
\end{eqnarray}

The modification in the radius due to the breakdown of Lorentz
symmetry appears in the term $6\alpha\omega\Bigl[1 -
\frac{12\beta}{\tau\Gamma}\Bigr]^{-3/2}$. If $\alpha =0$, then we
recover the usual prediction \cite{[27]}. If we denote
Chandrasekhar prediction by $R_{ch}$, then

\begin{eqnarray}
R = R_{ch}\Bigl\{1 - \frac{2}{5}\alpha\omega\Bigl[1 -
\Bigl(\frac{M}{\tilde{M}}\Bigr)^{2/3}\Bigr]^{-3/2}\Bigr\}
,\label{massradius2}
\end{eqnarray}

\begin{eqnarray}
\Bigl(\tilde{M}\Bigr)^{2/3} = \frac{\hbar c}{3\pi
G}\Bigl(\frac{9\pi}{8m_p}\Bigr)^{4/3}.\label{Constant5}
\end{eqnarray}

If $\alpha >0$, then the allowed radii are smaller than the
corresponding value when Lorentz symmetry is present, whereas
$\alpha <0$ yields larger radii. In order to see the possibilities
that our approach provides let us remember that \cite{[27]}

\begin{eqnarray}
R_{ch} =
\frac{(9\pi)^{1/3}}{2}\frac{\hbar}{mc}\Bigl(\frac{M}{m_p}\Bigr)^{1/3}\sqrt{1
- \Bigl(\frac{M}{\tilde{M}}\Bigr)^{2/3}}.\label{Chandradius1}
\end{eqnarray}

This last expression entails that if $M\rightarrow\tilde{M}$ (from
below), then $R_{ch}\rightarrow 0$. Let us now analyze the
consequences of (\ref{massradius2}). If $\alpha <0$, then in the
limit $M\rightarrow\tilde{M}$ the breakdown of Lorentz symmetry
predicts a non--vanishing radius for the white dwarf

\begin{eqnarray}
R \rightarrow -\frac{2}{5}\alpha\omega\Gamma\Bigl\{1 -
\Bigl(\frac{M}{\tilde{M}}\Bigr)^{2/3}\Bigr\}^{-1}
.\label{massradius3}
\end{eqnarray}

In other words, we have found for the case $\alpha>0$ a criterion
that could allow us to test this kind of violations to Lorentz
symmetry. Indeed, it predicts very large radii for white dwarfs
with mass very close to $\tilde{M}\sim 1.44M_s$, where $M_s$
denotes the mass of the sun. Those cases in which $\alpha
>0$ yield a smaller radius than Chandrasekhar's prediction.

\subsection{Case n=2}

In this case the pressure becomes

\begin{eqnarray}
P_0 = \frac{8\pi}{(2\pi \hbar)^3}(m^4c^5)\Bigl\{\frac{x^4_F -
x^2_F}{12} -
\frac{10}{105}\alpha\omega^2x^6_F\Bigr\}.\label{Groundpressure2}
\end{eqnarray}

The equilibrium condition for the white dwarf reads

\begin{eqnarray}
\frac{GM^2}{4\pi R^4} = \frac{8\pi}{(2\pi
\hbar)^3}(m^4c^5)\Bigl\{\frac{x^4_F - x^2_F}{12} -
\frac{10}{105}\alpha\omega^2x^6_F\Bigr\}.\label{Condition1n=2}
\end{eqnarray}

This last equation defines a dependence of the radius $R$ in terms
of the mass of the white dwarf $M$.

\begin{eqnarray}
R = R_{ch}\Bigl\{1 - \frac{4}{7}\alpha\omega^2\Bigl[1 -
\Bigl(\frac{M}{\tilde{M}}\Bigr)^{2/3}\Bigr]^{-2}\Bigr\}.\label{Condition2n=2}
\end{eqnarray}

In this last expression $\Gamma$ and $R_{ch}$ are given by
(\ref{Constant1}) and (\ref{Chandradius1}), respectively. Imposing
the condition $\alpha =0$ allows us to recover from
(\ref{Condition2n=2}) Chandrasekhar's relationship. Once again,
the present model predicts a non--vanishing radius for the case
$\alpha <0$ in the limit $M\rightarrow\tilde{M}$, though in this
case

\begin{eqnarray}
R \rightarrow -\frac{4}{7}\alpha\omega^2\Gamma\Bigl\{1 -
\Bigl(\frac{M}{\tilde{M}}\Bigr)^{2/3}\Bigr\}^{-3/2}.\label{massradius4}
\end{eqnarray}

If we divide (\ref{massradius4}) by (\ref{massradius3}) we find

\begin{eqnarray}
R_2/R_1 = \frac{10}{7}\omega\Bigl\{1 -
\Bigl(\frac{M}{\tilde{M}}\Bigr)^{2/3}\Bigr\}^{-1/2}.\label{Ratio1}
\end{eqnarray}

This last expression means that the limit $M\rightarrow\tilde{M}$
(for the case $\alpha<0$) diverges faster for $n=2$ than for
$n=1$. In other words, the effects of the breakdown of Lorentz
symmetry become more detectable as $n$ goes from 1 to 2.

\subsection{General Case}

The general equilibrium condition for a white dwarf is given by
the following expression

\begin{eqnarray}
\frac{GM^2}{4\pi R^4} = \frac{8\pi}{(2\pi
\hbar)^3}(m^4c^5)\Bigl\{\frac{x^4_F - x^2_F}{12} -
\frac{\alpha}{2}\omega^n \Bigl[\frac{2}{15}x^5_F\Bigl(1 +
x^2_F\Bigr)^{(n-1)/2} + \frac{n-1}{35}x^7_F\Bigl(1 +
x^2_F\Bigr)^{(n-3)/2}\Bigr]\Bigr\}.\label{Generalcondition1}
\end{eqnarray}

It is readily seen that the case $\alpha <0$ entails a pressure
larger than the corresponding value when Lorentz symmetry is
present. In other words, the breakdown of Lorentz symmetry in the
form of $\alpha <0$ can always be interpreted as the emergence of
a repulsive interaction among the particles of an electronic gas.
The question concerning the behavior of the radius in the limit
$M\rightarrow\tilde{M}$ for any value of $n$ requires a careful
analysis, though we may conjecture that it could always embody a
divergent radius, i.e., for any value of $n$ it seems that the
radius becomes ($a$ and $l$ are positive constants depending upon
the value of $n$)

\begin{eqnarray}
R = R_{ch}\Bigl\{1 - a\alpha\omega^n\Bigl[1 -
\Bigl(\frac{M}{\tilde{M}}\Bigr)^{2/3}\Bigr]^{-l}\Bigr\}.\label{Condition2n=3}
\end{eqnarray}

Hence in the limit $M\rightarrow\tilde{M}$, once again, we obtain
a divergent radius.

Finally, in the same line of reasoning, the case $\alpha >0$
implies for a fermionic gas the emergence of a lower pressure, and
therefore, the breakdown of Lorentz symmetry could always be
interpreted for this kind of matter equivalent to the emergence of
an attractive interaction among the particles of the gas.

\section{Conclusions}

In the present work a deformed dispersion relation has been
introduced as a fundamental fact for the dynamics of massive
bosons and fermions. The effects of this assumption upon the
thermodynamics of the corresponding gas have been analyzed.

For the case of massive bosons it has been proved that $\alpha >0$
is tantamount to the emergence of a repulsive interaction among
the particles, whereas, $\alpha <0$ is related to the appearance
of an attractive interaction. In other words, the breakdown of
Lorentz symmetry does impinge upon the thermodynamic properties of
a bosonic gas, entropy, state equation, specific heat, etc.,
though the possibility of detecting them is hindered by the fact
that the extra terms related to the loss of the symmetry behave
like $T/T_P$, where $T$ is the temperature of the system and
$T_p\sim10^{32}K$ is Planck's temperature. It has also been argued
that for those violations related to the breakdown of time
reversal invariance we may find that the entropy defines curves
without physical meaning, and that for these cases there are no
reversible adiabatic processes. In connection with this last
remark we confront two possibilities: (i) either thermodynamics
loses its validity within this context; (ii) or thermodynamics is
valid and those cases entailing the loss of time reversal
invariance have to be discarded.

At this point it is noteworthy to comment that in the extant
literature we may find some results claiming that the case $n=1$
has to be discarded due to violations to black hole thermodynamics
(\cite{[28], [29]}). The present results support the conclusions
contained in (\cite{[28], [29]}), since they coincide in the fact
that the linear case has problems with the concept of entropy. The
new point in our work consists in the fact that these problems
with thermodynamics can be found also in the context of ordinary
matter, for instance, a bosonic gas.

Additionally, the case of fermionic statistics has been analyzed,
and in the quest for a system that could provide a feasible
experimental proposal for the detection of this kind of violations
of Lorentz symmetry the modifications that the Chandrasekhar's
mass--radius relationship suffers have been considered. This allow
us to introduce astrophysical objects and try to understand if
they could shed some light upon this issue. For the case of
massive fermions it has been proved that $\alpha <0$ is equivalent
to the emergence of a repulsive interaction among the particles,
whereas, $\alpha
>0$ is related to the appearance of an attractive interaction.
Additionally, it has been proved that the case $\alpha <0$
embodies a behavior (in the limit $M\rightarrow\tilde{M}$) very
different from the predictions of the usual model for white
dwarfs. The current data seems to discard negative values of
$\alpha$ \cite{[30]}. In other words, our approach, together with
the current data, could allow us to consider only positive values
of $\alpha$ as physically meaningful. At this point it is
noteworthy to comment that more observations are required to check
the prediction of the model for the case $\alpha <0$, since the
closest value to our critical mass, $\tilde{M}\sim1.44M_s$ (for
Sirius B, $M/M_s = 1.0034\pm 0.026$ \cite{[30]}) lies not close
enough to the needed value.

It is interesting to mention that the observations contain a
puzzling feature, namely, some stars do have radii which are
significantly smaller than the theoretical predictions \cite{[19],
[30]}. There are several models which (without resorting to any
kind of breakdown of Lorentz symmetry) try to solve this puzzle.
Nevertheless, at this point we must mention that these
observations are compatible with $\alpha
>0$, though not necessarily provide a proof for the existence of
non--vanishing values of $\alpha$. Let us explore the
possibilities that these observations could mean in this context.
and denote the difference between observation and Chandrasekhar's
model in the radius by $\Delta R$, for $\alpha
>0$. Then, for $n=1$

\begin{eqnarray}
\frac{5}{2\omega\Gamma}\Delta
R\Bigl[1-\Bigl(\frac{M}{\tilde{M}}\Bigr)^{2/3}\Bigr]\geq
\alpha.\label{massradius6}
\end{eqnarray}

Let us now consider the following two white dwarfs with the same
mass, i.e., G156--64 (a strange white dwarf) and Wolf 485 A (this
white dwarf satisfies Chandrasekhar's relationship), see table I
in \cite{[19]}. In this case $\Delta R =2.78\times 10^{8}$cm, and
therefore $\alpha\leq 10^{22}$. Taking into account further
physical aspects, for instance, Coulomb correction, lattice
energy, or more realistic density distribution for the electronic
gas, etc., shall provide a much lower bound for $\alpha$.

Finally, the present approach could also be implemented in
connection with other schemes, for instance, $\kappa$--Poincar\'e
dispersion relation in order to obtain constraints upon the
quantum $\kappa$--Poincar\'e algebra. This possibility would be,
in some sense, a continuation of work already done \cite{[31]}.

\begin{acknowledgments}
 This research was supported by CONACYT Grant 47000--F. The author would like to thank
 A. A. Cuevas--Sosa and A. Mac\'{\i}as for useful discussions and literature hints.
\end{acknowledgments}

\end{document}